\documentclass[a4paper,11pt]{article}
\usepackage{fullpage}
\usepackage{amsmath}
\usepackage{color}

\usepackage{amsthm} 
\usepackage{amssymb} 
\usepackage{graphicx}

\usepackage[position=top]{subfig}
\usepackage{bbold}
\usepackage{stmaryrd} 

\usepackage[square, sort, numbers]{natbib} 
\usepackage[colorlinks,urlcolor=red,citecolor=red,linkcolor=blue]{hyperref}

\title{Numerical Methods for the Stray-Field Calculation:\\ A Comparison of recently developed Algorithms}

\author{Claas Abert\thanks{Funded by the Deutsche Forschungsgemeinschaft via the Graduiertenkolleg 1286 ``Functional Metal-Semiconductor Hybrid Systems''}\\ Fachbereich Mathematik, Universit\"at Hamburg,\\ Bundesstr. 55, 20146 Hamburg, Germany\\[0.1cm] email: \texttt{cabert@physnet.uni-hamburg.de},\\[0.3cm]
Lukas Exl\thanks{Funded by the Austrian Science Fund (FWF, project SFB F4112-N13)}\\ University of Applied Sciences, Department of Technology\\ A-3100 St.Poelten, Austria\\[0.1cm] email: \texttt{lukas.exl@fhstp.ac.at}
\\[0.3cm] --$\,\Diamond\,$--\\[0.3cm]
Gunnar Selke\thanks{Funded by the Deutsche Forschungsgemeinschaft via the Graduiertenkolleg 1286 ``Functional Metal-Semiconductor Hybrid Systems''}\\ Arbeitbereich Technische Informatiksysteme, Universit\"at Hamburg,\\Vogt-K\"olln-Str. 30, 22572 Hamburg, Germany\\[0.1cm]
Andr\'e Drews\thanks{Funded by the Sonderforschungsbereich 668 ``Magnetism from the single atom to the nanostructure''}\\ Arbeitbereich Technische Informatiksysteme, Universit\"at Hamburg,\\Vogt-K\"olln-Str. 30, 22572 Hamburg, Germany\\[0.1cm]
Thomas Schrefl\\ University of Applied Sciences, Department of Technology\\ A-3100 St.Poelten, Austria
}

\begin{document}
\maketitle

\begin{abstract}
Different numerical approaches for the stray-field calculation in the context of micromagnetic simulations are investigated.
We compare \textit{finite difference based fast Fourier transform methods}, \textit{tensor grid methods} and the \textit{finite-element method with shell transformation} in terms of computational complexity, storage requirements and accuracy tested on several benchmark problems.
These methods can be subdivided into integral methods (fast Fourier transform methods, tensor-grid method) which solve the stray field directly and in differential equation methods (finite-element method), which compute the stray field as the solution of a partial differential equation.
It turns out that for cuboid structures the integral methods, which work on cuboid grids (fast Fourier transform methods and tensor grid methods) outperform the finite-element method in terms of the ratio of computational effort to accuracy.
Among these three methods the tensor grid method is the fastest.
However, the use of the tensor grid method in the context of full micromagnetic codes is not well investigated yet.
The finite-element method performs best for computations on curved structures.\\[0.3cm]
\end{abstract}
  
\section{Introduction}
Micromagnetic simulations nowadays are highly important for the investigation of ferromagnetic materials which are used in storage systems and electric motors and generators.
In these simulations the magnetic state of the ferromagnet is represented by a classical magnetization vector field.

The computation of the non-local magnetostatic interactions is thereby the most time-consuming part.
Naive implementation of the superposition-based integral operators \eqref{eq:intop} or solvers for the underlying differential equation (Poisson equation \eqref{eq:poisson}) yield computational costs proportional to the square of the number of grid points, i.e. $\mathcal{O}(N^2)$.
Several methods have been introduced in literature in order to reduce these costs.

The magnetic scalar potential can be computed by solving the Poisson equation.
The solution of the Poisson equation with the finite-element method (FEM) has a complexity of $\mathcal{O}(N)$ if boundary conditions are given at the boundary of the sample and a multigrid preconditioner is used \cite{tsukerman_1998}.
However, the stray-field problem has open boundary conditions, where the potential is known at infinity.
Two possible solutions for the open boundary problem are the coupling of the boundary element method (BEM) with the finite-element method \cite{fredkin_1990} and the application of a shell transformation \cite{brunotte_1992}.
BEM gives an additional complexity of $\mathcal{O}(M^2)$ where $M$ is the number of boundary nodes.
This complexity can be reduced to $\mathcal{O}(M \log M)$ by application of the $\mathcal{H}$-matrix approximation for the dense and unstructured boundary element matrices \cite{buchau_2003,pop_2004,knittel_2009}.
The storage requirements and computational complexity of the FEM with shell transformation will be described in the forthcoming text.

Another class of methods rely on the evaluation of volume and/or surface integrals for the direct computation of the magnetostatic potential or the field, e.g. fast multipole methods \cite{FMM_1997,blue_1991}, nonuniform grid methods \cite{NG_2009} and fast Fourier transform (FFT) methods \cite{abert_2012,long_2006}, scaling from $\mathcal{O}(N)$ to $\mathcal{O}(N\log N)$. The more recent tensor grid method (TG), which also belongs to this class scales even better under certain assumptions.

In this paper we compare recently developed algorithms, namely the FFT-based methods for the computation of the field via the scalar potential (SP) and directly (DM) \cite{abert_2012,magnum_homepage,magnum_2012}, a recently developed approach from numerical tensor-structered methods (TG) \cite{exl_fast_2012}, and the finite-element method with shell transformation (FES), which is a FEM method that does not rely on BEM approaches and thus only introduces sparse matrices.

\section{Stray-Field Problem}
Consider a magnetization configuration $\boldsymbol{M}$ that is defined on a finite region $\Omega = \{\boldsymbol{r} : \boldsymbol{M}(\boldsymbol{r}) \neq 0\}$.
In order to perform minimization of the full micromagnetic energy functional or solve the Landau-Lifshitz-Gilbert (LLG) equation it is necessary to compute the stray field within the finite region $\Omega$.
The stray-field energy is given by
\begin{align}
  e_d = -M_\text{s} \frac{1}{2} \int_\Omega \boldsymbol{M} \cdot \boldsymbol{H} \;\text{d}^3 \,\boldsymbol{r}.
\end{align}
The Landau-Lifshitz-Gilbert equation reads
\begin{align}
  \boldsymbol{M}_t = - \frac{\gamma}{1+\alpha^2} \boldsymbol{M} \times \boldsymbol{H}_{\text{eff}} + \frac{\alpha \gamma}{M_\text{s} (1 + \alpha^2)} \boldsymbol{M} \times (\boldsymbol{M} \times \boldsymbol{H}_{\text{eff}}),
\end{align}
where $\alpha$ is the Gilbert damping constant and $\boldsymbol{H}_\text{eff}$ is the effective field given by the negative variational derivative of the energy density.
In both cases the stray field is only required to be known within $\Omega$.
The stray field $\boldsymbol{H}$ has a scalar potential $\phi$, which is the solution of a Poisson equation \cite{jackson_classical_1999}
\begin{align}\label{eq:poisson}
  \boldsymbol{H} &= -\nabla \phi \\
  \Delta \phi    &= \nabla \cdot \boldsymbol{M}
\end{align}
The stray field $\boldsymbol{H}$ and thus also the scalar potential $\phi$ are required to vanish at infinity.
This boundary condition is often referred to as open boundary condition \cite{fidler_2000}.

\section{Methods}
\subsection{FFT Methods (SP and DM)}
One way to reduce the computational complexity is to employ the fast Fourier transform (FFT).
FFT methods solve an integral solution of the Poisson equation by applying the convolution theorem.
The solution to the Poisson equation \eqref{eq:poisson} is given by the integral, see \cite{jackson_classical_1999},
\begin{align}\label{eq:intop}
  \phi(\boldsymbol{r}) =
  - \frac{1}{4\pi} \int_\Omega \frac{\nabla' \cdot \boldsymbol{M}(\boldsymbol{r'})}{|\boldsymbol{r} - \boldsymbol{r'}|} \text{d}^3 \boldsymbol{r'}
  + \frac{1}{4\pi} \int_{\partial \Omega} \frac{\boldsymbol{n'} \cdot \boldsymbol{M}(\boldsymbol{r'})}{|\boldsymbol{r} - \boldsymbol{r'}|} \text{d} A'
\end{align}
which directly fulfills the required open boundary condition.
Performing integration by parts yields
\begin{align}\label{scpot}
  \phi(\boldsymbol{r})
  &= \frac{1}{4\pi} \int_\Omega \boldsymbol{M}(\boldsymbol{r}') \cdot \boldsymbol{\nabla'} \frac{1}{|\boldsymbol{r} - \boldsymbol{r'}|} \text{d}^3 \boldsymbol{r'} \\
  &= \boldsymbol{S}(\boldsymbol{r} - \boldsymbol{r'}) \ast \boldsymbol{M}(\boldsymbol{r'}).
\end{align}
By employing the convolution theorem
\begin{align}\label{convolution}
  \phi
  = \boldsymbol{S} \ast \boldsymbol{M}
  = \mathcal{F}^{-1} \Big( \mathcal{F}(\boldsymbol{S}) \cdot \mathcal{F}(\boldsymbol{M}) \Big),
\end{align}
this convolution can be discretized and solved with the fast Fourier transform.
A prerequisite for this procedure is the usage of an equidistant grid, which is required for a discrete convolution.
The stray field
\begin{align}\label{field}
 \boldsymbol{H}(\boldsymbol{r}) = -\boldsymbol{\nabla}\,\boldsymbol{\phi}(\boldsymbol{r}), 
\end{align}
can be obtained by applying finite differences.
This method is referred to as the scalar-potential method (SP) in the following.
It is described in detail in \cite{abert_2012}.

It is also possible to compute the stray field $\boldsymbol{H}$ directly as a result of a matrix--vector convolution.
\begin{align}\label{demag}
  \boldsymbol{H}(\boldsymbol{r})
  &= - \frac{1}{4\pi} \int \Big(\boldsymbol{\nabla} \boldsymbol{\nabla}' \frac{1}{|\boldsymbol{r} - \boldsymbol{r'}|}\Big) \boldsymbol{M}(\boldsymbol{r}')\text{d}^3 \boldsymbol{r'} \\
  &= \boldsymbol{N}(\boldsymbol{r} - \boldsymbol{r'}) \ast \boldsymbol{M}(\boldsymbol{r'}).
\end{align}
Here $\boldsymbol{N}$ denotes the demagnetization tensor.
Similar to \eqref{convolution} the convolution can be solved as an element-wise matrix--vector multiplication in Fourier space.
This method is referred to as the demagnetization-tensor method (DM) in the following and is implemented by different finite-difference codes \cite{oommf,arne_2011,magnum_2012}.
For the numerical experiments we use MicroMagnum \cite{magnum_homepage,magnum_2012} which implements both the SP and the DM method.
\subsection{Tensor Grid Methods (TG)}
Tensor grid methods (TG) for micromagnetic stray-field computation were recently introduced in \cite{exl_fast_2012} and \cite{goncharov_2010}. They were developed for the purpose of handling so called low-rank tensor or compressed tensor magnetization, see \cite{kolda_tensor_2009} for a survey, in order to accelerate the computations and relieve storage requirements, see \cite{exl_micro_2012}. In the following we give a brief introduction into the ideas behind this method, also see \cite{exl_fast_2012} for a detailed description.

\subsubsection{Analytical preparations}  
The computation of the stray field within the magnetic body is based on the explicit integral formula for the scalar potential \eqref{scpot}.
The main idea is the usage of a representation for the integral kernel in \eqref{scpot} as an integral of a \textit{Gaussian function} by the formula
\begin{align}\label{transform}
\frac{1}{|\boldsymbol{r}-\boldsymbol{r}^{\,\prime}|^{3}} = \frac{2}{\sqrt{\pi}} \int_{\mathbb{R}} \tau^2\,e^{-\tau^2 |\boldsymbol{r}-\boldsymbol{r}^{\,\prime}|^{2}}\,\text{d}\tau,
\end{align}
which leads from \eqref{scpot} to
\begin{align}\label{gausspot}
\phi(\boldsymbol{r}) = \frac{1}{2\pi^{\frac{3}{2}}} \int_{\mathbb{R}} \tau^2 \int_{\Omega}
                 e^{-\tau^2\,|\boldsymbol{r}-\boldsymbol{r}^{\,\prime}|^2}\boldsymbol{M}(\boldsymbol{r}^{\,\prime})\cdot (\boldsymbol{r} - {\boldsymbol{r}^{\,\prime}})
                \,\text{d}^{\,3} \boldsymbol{r}^{\,\prime}\,\text{d}\tau.
\end{align}
Equation \eqref{gausspot} reduces the computation to independent spatial integrals along each principal direction (the part of the $\Omega$ integral without the magnetization is now a product of independent $1$--D integrals). This analytical preparation directly results in a reduction of the computational effort from $\mathcal{O}(N^2)$ to $\mathcal{O}(N^{4/3})$ if discretized on a \textit{tensor product grid} before even using compressed/low-rank tensor formats for the discretized magnetization components. A similar method was introduced for the computation of the electrostatic scalar potential, \cite{juselius_2007}.\\
The additional $\tau$-integration is carried out by the exponentially convergent \textit{Sinc quadrature} \cite{hackbusch_low-rank_2005}, the spatial integrals are computed by \textit{Gauss-Legendre quadrature}, both resulting in a numerical error of about the machine epsilon.\\

\begin{figure}
  \centering
  \includegraphics{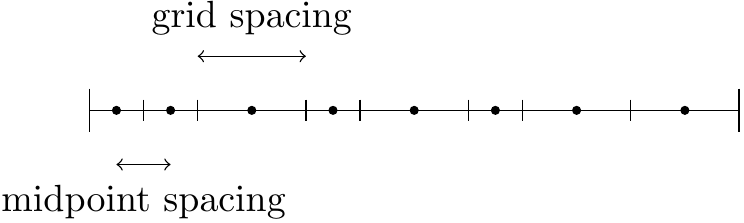}
  \caption{Grid spacing and midpoint spacing in TG Methods}
  \label{fig:tensor_spacing}
\end{figure}

\subsubsection{Discretization on a tensor product grid}
The magnetic body $\Omega$ is discretized on a \textit{tensor product grid} arising from the tensor outer product of three vectors $h_p \in \mathbb{R}^{N_p}, p = 1\hdots 3$ related to the grid spacings along each axis (see Fig.~\ref{fig:tensor_spacing}).  This results in a not necessarily uniform \textit{Cartesian grid} but in contrast to methods like DM/SP described before, tensor grid methods make use of the tensor-product interpretation of such grids.\\ 
The magnetization on the center points of the cells is given as a $3$-tensor \cite{kolda_tensor_2009} for each component, i.e.
\begin{align}
  \boldsymbol{\mathcal{M}}^{(p)} \in \mathbb{R}^{N_1 \times N_2 \times N_3}, \quad p=1\hdots 3
\end{align}   
where $N_1,N_2,N_3$ are the number of cells in the principal directions.
Thereby it is possible to use \textit{low-rank} representation for the magnetization like \textit{Canonical/Parallel Factors Decomposition} (CP) or \textit{Tucker formats}, see the Appendix~\ref{appendix} or \cite{kolda_tensor_2009}.
We obtain the potential on the center points of the computational cells, as the discrete analogue of \eqref{gausspot}, by a so-called \textit{block-CP tensor} \cite{exl_fast_2012} 
\begin{align}\label{sumpot}
\mathbb{R}^{N_1 \times N_2 \times N_3} \ni \Phi = \frac{1}{2 \pi^{3/2}} \sum_{p=1}^{3} \sum_{l=1}^{R} \omega_{l}\,\sinh(\tau_{l})^2\,\boldsymbol{\mathcal{M}}^{(p)} \times_1 \boldsymbol{D}_{1}^{\,l} \times_2 \boldsymbol{D}_{2}^{\,l} \times_3 \boldsymbol{D}_{3}^{\,l}. 
\end{align}
Here $(\tau_l , \omega_l )$ are the nodes and weights arising from the Sinc-quadrature with $R$ terms. 
The \textit{Gaussian matrices} $\boldsymbol{D}_{q}^{\,l} \in \mathbb{R}^{N_q \times N_q}$ come from
\begin{align} \label{Dmat}
d_{i_q,j_q}^{\,l} := & \int_{\Omega_{j_{q}}} g(x_{i_{q}}^{c}, x^{\prime}, \tau_{l})\,dx^{\prime}, \\
\boldsymbol{D}_{q}^{\,l} := & \big( d_{i_q\,j_q}^{\,l} \big).
\end{align}
where $\Omega_{j_{q}}$ denotes the $j$-th interval on the (partitioned) $q$-th axis with length $(h_q)_j$ and $x_{i_{q}}^{c}$ is the midpoint of the $i$-th interval on the $q$-th axis. The Gaussian integrands are given by 
\begin{align}\label{gs}
g^{(q)}(\alpha,\alpha^{\prime}, \tau) :=
    \left\{\begin{array}{l l}
     \exp(-\sinh(\tau)^2\,(\alpha - \alpha^{\prime})^2)                              & q \neq p, \\*[\jot]
     (\alpha - \alpha^{\prime}) \exp(-\sinh(\tau)^2\,(\alpha - \alpha^{\prime} )^2)  & q = p,
  \end{array}\right.
\end{align}
and are approximated using Gauss-Legendre quadrature, 
as mentioned above.
The field within $\Omega$ is derived from \eqref{field} by finite-difference operators of second order.

\subsubsection{Low-Rank Magnetization}
Equation \eqref{sumpot} allows the treatment of specially structured magnetization tensors, like CP or Tucker tensors \cite{kolda_tensor_2009} that have a reduced number of degrees of freedom, see Tab.~\ref{tab1}, and accelerate the computation up to \textit{sub-linear} effort (below the volume size $N^3$), see Tab.~\ref{tab2}. As a consequence TG methods using low-rank magnetization allow larger models with finer discretization density than conventional methods.\\

\begin{figure}
  \centering
  \subfloat[]{\includegraphics[width=0.4\textwidth]{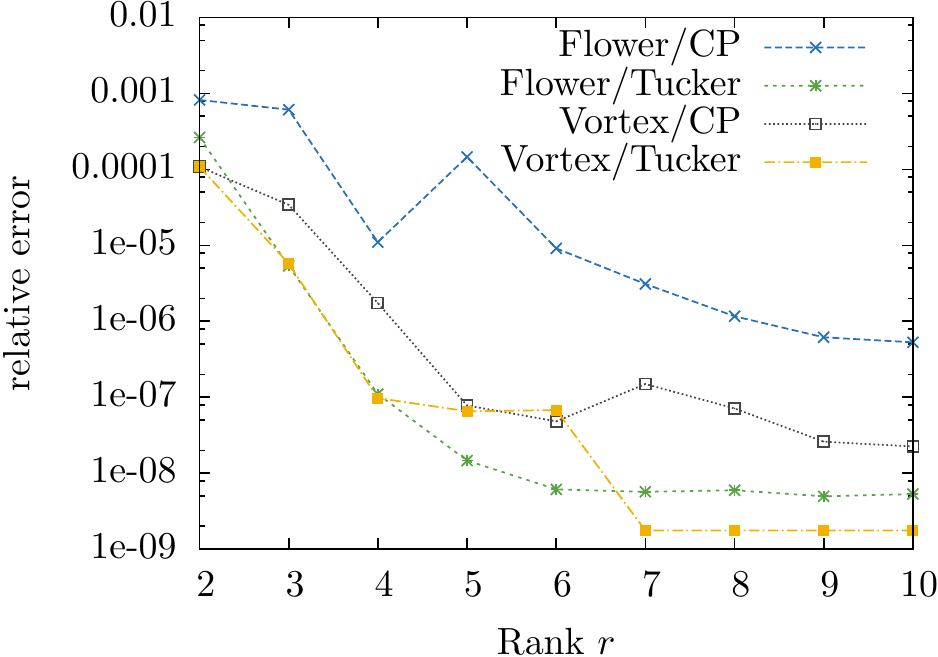}\label{fig:low_rank_r}}
  \hspace{1cm}
\subfloat[]{\includegraphics[width=0.4\textwidth]{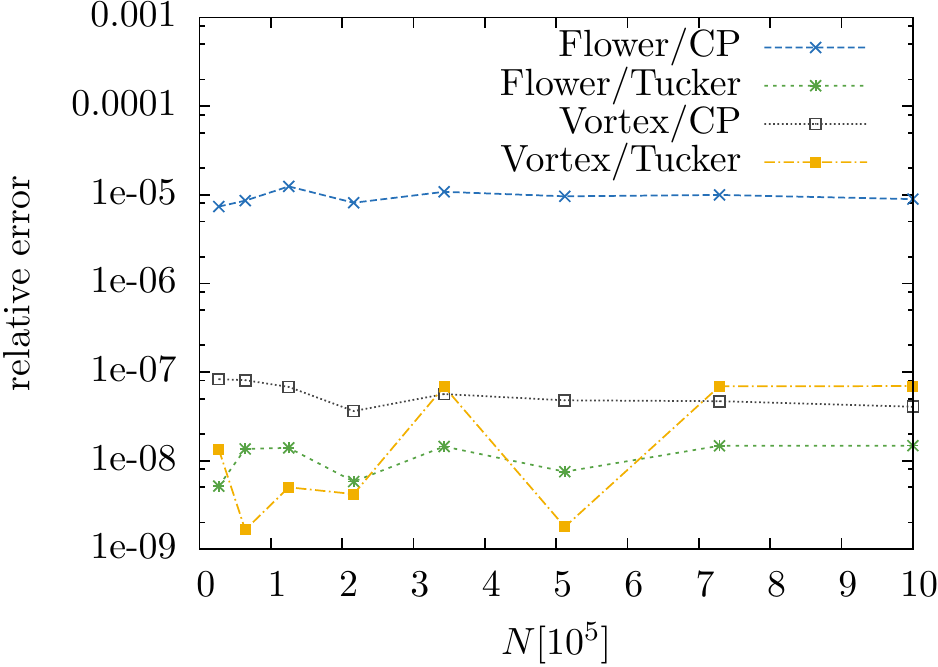}\label{fig:low_rank_n}}
  \caption{
    Low-rank approximation of flower and vortex state via Tucker and CP decomposition.
    (a) Relative error w.r.t. approximation rank $r$. $N=1e$+$06$.
    (b) Relative error w.r.t. discretization density $N$. Rank $r = 5$.
  }
  \label{fig:low_rank}
\end{figure}

We now show by means of numerical experiments that typical single domain states \cite{mumag3} have highly accurate low-rank representations.
Fig.~\ref{fig:low_rank} shows the approximation properties of a flower and a vortex state as described in Sec. \ref{exps} via the CP format and the Tucker format using an \textit{alternating least squares algorithm} (ALS) \cite{kolda_tensor_2009} for the approximations.
The plots \ref{fig:low_rank_r} and \ref{fig:low_rank_n} are computed independently from random initial guesses used in the ALS algorithm.
We set the parameters in \eqref{flower_mag} as $a=c=0.5$, $b=1$ and in \eqref{config1} as $r_c = 1/2$. 
Fig.~\ref{fig:low_rank_r} shows the dependence of the relative error \eqref{relr} w.r.t. the rank for fixed discretization density, where Fig.~\ref{fig:low_rank_n} indicates the dependence w.r.t. the discretization density $N$ for fixed rank.\\
The relative errors are measured in the Frobenius norm, i.e.
\begin{align}\label{relr}
relerr = \Big(\sum_{p=x,y,z} \left\|\boldsymbol{\mathcal{M}}_{\text{dense}}^{(p)} - \boldsymbol{\mathcal{M}}_{\text{low-rank}}^{(p)}\right\|_{F}^2  \Big)^{\frac{1}{2}}/\Big(\sum_{p=x,y,z} \left\|\boldsymbol{\mathcal{M}}_{\text{dense}}^{(p)}\right\|_{F}^2  \Big)^{\frac{1}{2}}.
\end{align}
The Tucker format generally leads to a better approximation, where Fig.~\ref{fig:low_rank_n} essentially shows no loss of accuracy while increasing the mesh density.\\ 
       
\subsection{FEM Methods (FES)}
Within the finite-element framework the Poisson equation \eqref{eq:poisson} is solved by the weak formulation
\begin{align}
  \int_{\Omega} \nabla \phi \cdot \nabla v \;\text{d}^3 x =
  \int_{\Omega} \boldsymbol{M} \cdot \nabla v \;\text{d}^3 x
  \quad \forall \quad v,\phi \in V
\end{align}
where Dirichlet boundary conditions are embedded in the trial function space $V$, i.e. the function space of the solution $\phi$.
In the case of the stray-field problem the boundary conditions at the sample boundary $\partial \Omega$ are unknown.
They are defined as zero at infinity.
The treatment of these open boundary conditions is the main difficulty for finite-element stray-field calculations.

\begin{figure}
  \centering
  \subfloat[]{\includegraphics[trim=200 0 200 0,clip,width=0.3\textwidth]{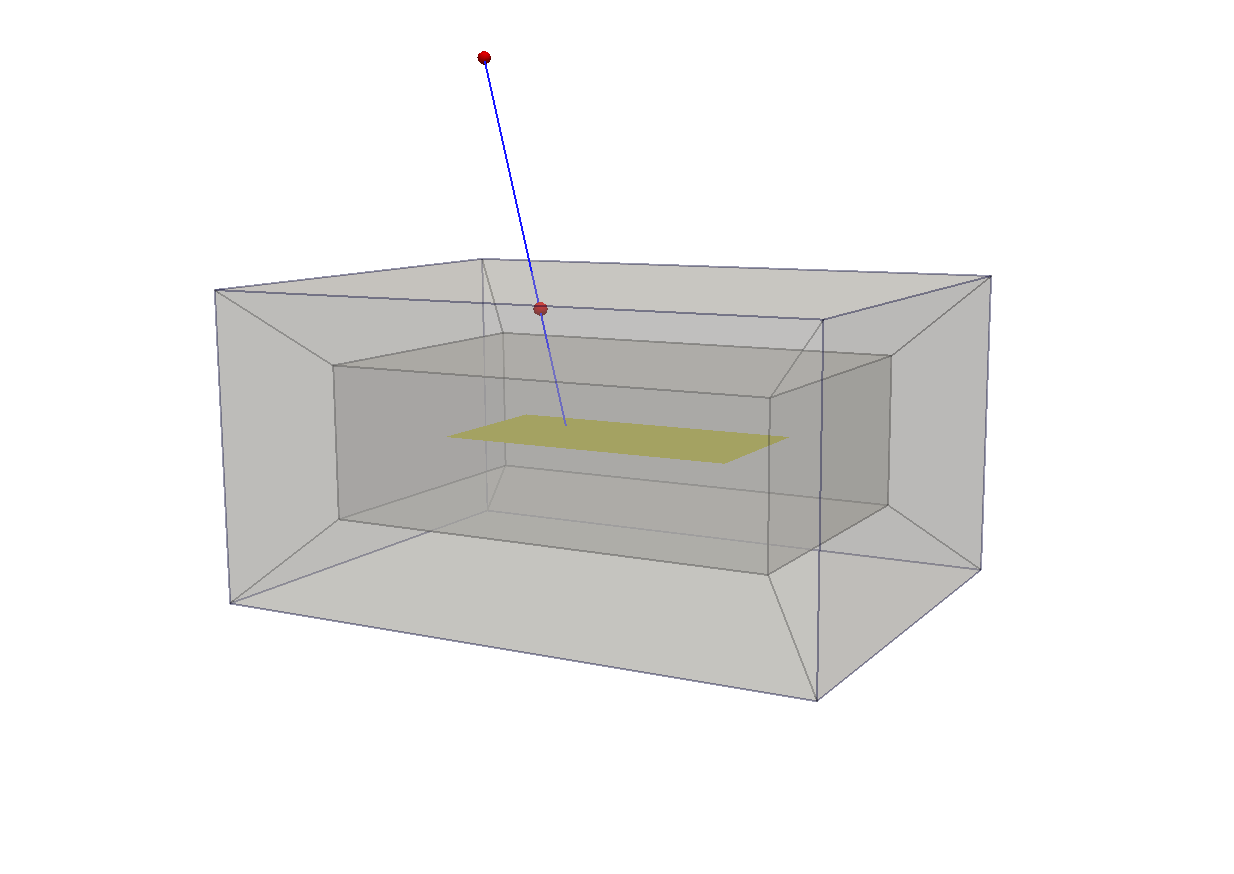}\label{fig:transformation_sketch}}
  \hspace{1cm}
\subfloat[]{\includegraphics[trim=200 0 200 0,clip,width=0.3\textwidth]{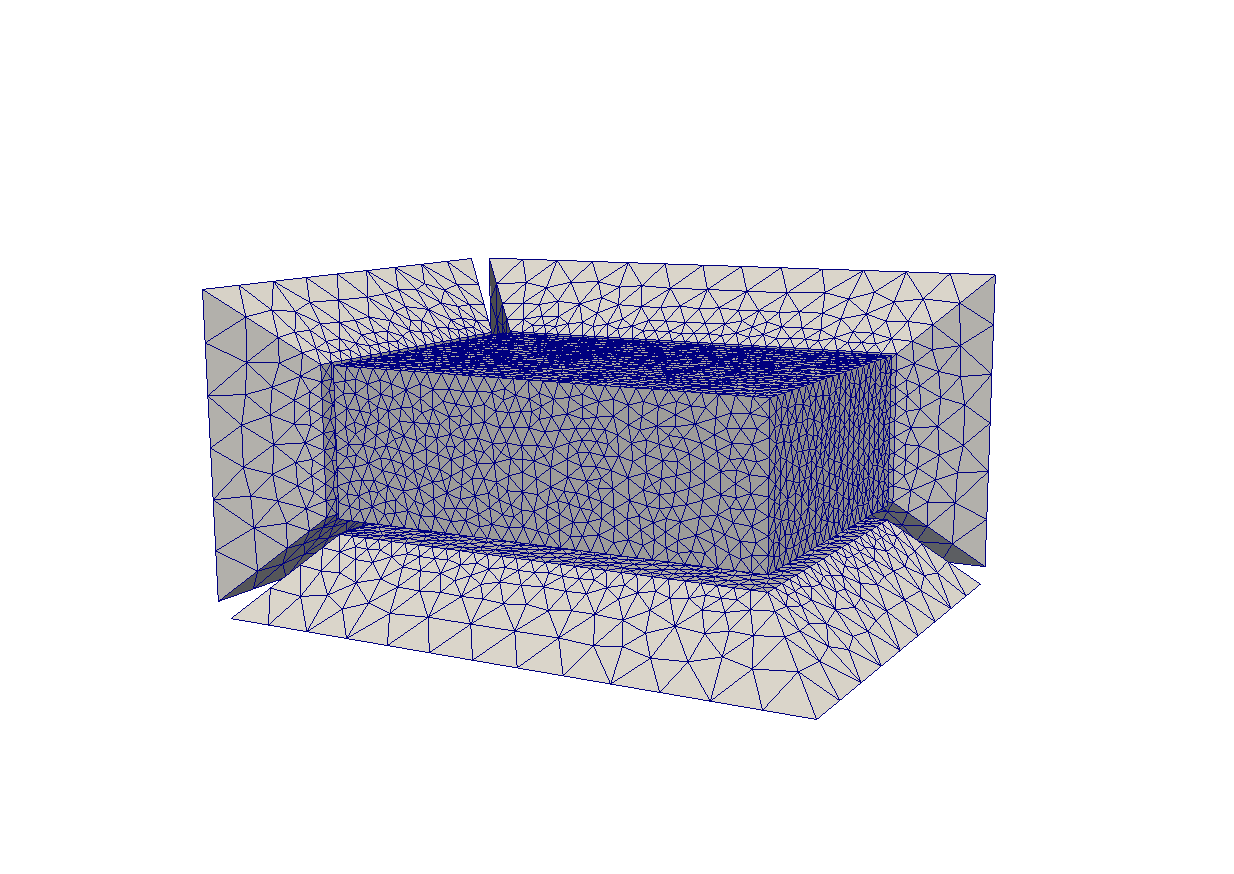}\label{fig:transformation_mesh}}
  \caption{
  Parallelepipedic shell surrounding the cuboid sample.
  (a) The transformation is carried out along the blue line.
      The origin of the transformation moves along the yellow middle plane.
  (b) Since the area of interest is the sample, the mesh is coarsened in the shell.
  }
  \label{fig:transformation}
\end{figure}
\begin{figure}
  \centering
  \includegraphics{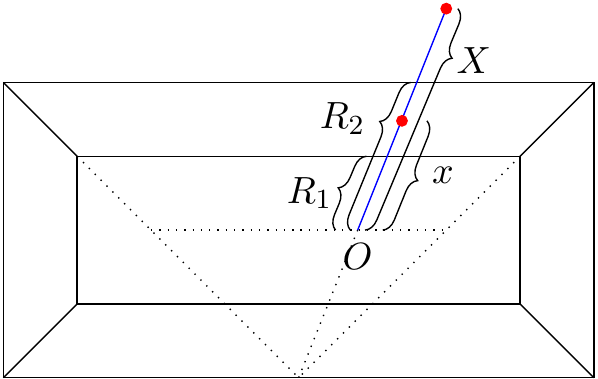}
  \caption{
    Sketch of the shell transformation in two dimensions.
    Within each shell patch the shell points are transformed in a radial sense.
    In order to achieve continuity between the patches the origin $O$ of the one-dimensional transformation has to be continuous between the patches.
    We choose the origin $O$ to move on the middle plane of the sample.
    The third dimension is treated in the same manner.
  }
  \label{fig:transformation2}
\end{figure}
We present the results for a transformation technique.
The sample is surrounded by a finite shell which is also meshed.
A bijective transformation from the finite shell to the complete exteriour of the sample is applied by introducing a metric tensor to the weak formulation.
The particular transformation we use is known as ``parallelepipedic shell transformation'' \cite{brunotte_1992}.
The sample is put into a cuboid volume and a shell consisting of six parallelepipeds is created (see Fig.~\ref{fig:transformation_sketch}).

The transformation is chosen such that points located at the inner boundary of the shell are mapped to themselves.
Points on the outer boundary of the shell are mapped to infinity.
The Jacobian of the transformation is requested to be 1 on the inner boundary of the shell in order to be continuous across the sample boundary.

These conditions still leave some space for the choice of transformation.
The most important aspect of this method is the distortion of the test and trial functions in the transformed area.
In order to get a good result, the test and trial functions must be distorted such that they are able to model the natural decay of the magnetic potential.
This obviously also depends on the choice of test and trial functions.
From \eqref{scpot} it is seen that the potential decays with $1/r^2$.
We choose our test and trial function $\phi_h, v_h \in V_h$ to be continuous and piecewise third-order polynomial ($\mathcal{P}_3)$
\begin{align}
  V_h = \Big\{ v_h \in H^1(\Omega):v_h|_T \in \mathcal{P}_3(T) \;\forall\; T \in \mathcal{T}_h \Big\}
\end{align}
where $H^1$ is a Sobolev space and $\mathcal{T}_h$ is a tetrehedron tesselation (see Fig.~\ref{fig:transformation_mesh}).
The transformation per shell patch is carried out in a radial sense as sketched in Fig.~\ref{fig:transformation2}.
The scalar transformation is given by
\begin{align}
  X = R_1 \frac{R_2 - R_1}{R_2 - |x|}
\end{align}
with $R_1$, $R_2$, $x$, and $X$ as shown in Fig. \ref{fig:transformation2}.
This transforms the third order polynomial test and trial functions as
\begin{align}
  a + bx + cx^2 + dx^3 \rightarrow a' + b'\frac{1}{X} + c'\frac{1}{X^2} + d'\frac{1}{X^3}.
\end{align}
The discretized weak formulation then reads
\begin{align}
  \int_\Omega (\nabla \phi_h)^T \boldsymbol{g} \nabla v_h \;\text{d}^3 x &= 
  \int_{\Omega} \boldsymbol{M} \cdot \nabla v_h \;\text{d}^3 x
  \quad \forall \quad v_h \in V_h\\
  \boldsymbol{g} &= \left\{
  \begin{array}{ll}
    \mathbb{1} & \text{if}\; \boldsymbol{x} \in \Omega_\text{sample}\\
    {\boldsymbol{J}^{-1}}^T \;|\boldsymbol{J}|\; \boldsymbol{J}^{-1} & \text{if}\; \boldsymbol{x} \in \Omega_\text{shell}
  \end{array} \right.
\end{align}
where $\Omega_\text{sample}$ and $\Omega_\text{shell}$ denote the disjoint regions of the sample and the transformed shell with $\Omega_\text{sample} \cup \Omega_\text{shell} = \Omega$ and  $\boldsymbol{J}$ is the Jacobian matrix of the transformation.
This directly translates to a linear system of equations, where the solution vector contains the coefficients in terms of the discrete function basis.
The size of this solution vector is referred to as degrees of freedom (DoF).
The implementation of this method is done with FEniCS \cite{fenics_book}.
\section{Storage Requirements and Computational Complexity}\label{storage}
The costs of the different methods are compared in terms of storage requirements and computational complexity.
Table \ref{tab1} and \ref{tab2} show the results.
We choose $N$ to be the number of computational cells in the case of 
DM, SP and TG methods.
In the case of finite-element methods (FES) $N$ refers to the number of degrees of freedom.

Besides the memory needed for the storage of the magnetization configuration and the stray field, all methods require a certain amount of extra storage for auxiliary constants.
In case of the DM and SP methods this includes the convolutions kernels, TG needs the one dimensional Gaussian matrices, and finite-element methods (FES) require the stiffness matrix as an auxiliary constant.
These constants depend on the geometry and discretization only.
This means that the computation of auxiliary constants has to be done only once for different magnetization configurations.
Thus their complexity is almost irrelevant in the context of LLG computations and energy minimization.
The storage requirements for these constants as well as the computational complexity of their calculation is summarized in the setup column in both tables.

Storage requirements for TG methods depend on the rank $r$ used for the low-rank tensor representation of the magnetization components. Often the rank is much smaller than $N^{1/3}$, the discretization size in one spatial dimension. This makes the storage requirements for the magnetization, potential and field proportional to $rN^{1/3}$. For the setup the ($N^{1/3}\times N^{1/3}$) Gaussian matrices need to be computed and stored, thereby $R$ in Tab.~\ref{tab1} and \ref{tab2} denotes the number of Sinc-quadrature nodes.
The computational effort in TG methods also depends on the tensor format used for the representation of the magnetization, see Tab.~\ref{tab2}. If the magnetization has a low-rank representation, TG methods usually reduce this complexity below the number of computational cells (\textit{sub-linear}), making this methods the fastest available nowadays.
 
The storage requirements for the other three methods are proportional to $N$, which is a result of the dense representation of the magnetization, see Tab.~\ref{tab1}.
A well-known result is the $N\log{N}$ complexity of the convolution in FFT methods (DM/SP), likewise this is the asymptotic operation count for those methods, Tab.~\ref{tab2}.
In the FES method sparse linear systems have to be solved for the computation of the scalar potential.
We used a conjugate gradient solver (CG) with an algebraic multigrid preconditioner (AMG) and measured the complexity w.r.t. $N$ (DoF) experimentally, finding a linear dependence on the system size (with a small logarithmic scaling factor). 

\begin{table}
  \centering
  \begin{tabular}{l c c c c }
    Method  &  Setup & Magnetization & Potential & Field \\ \hline\hline
    DM          &  $48N$          & $3N$               & --                 & $3N$ \\ \hline
    SP          &  $24N$          & $3N$               & $N$                & $3N$ \\ \hline
    TG (dense)  &  $6RN^{2/3}$    & $3N$               & $N$                & $3N$ \\ \hline 
    TG (Tucker) &  $6RN^{2/3}$    & $3(r^3+3rN^{1/3})$  & $9R(r^3+3rN^{1/3})$ & $27R(r^3+3rN^{1/3})$ \\ \hline
    TG (CP)     &  $6RN^{2/3}$    & $3(r+3rN^{2/3})$        & $9(r+3rRN^{1/3})$       & $27(r+3rRN^{1/3})$  \\ \hline
    FES         &  $\approx 48N$  & $3N$               & $N$                & $3N$ \\ \hline
  \end{tabular}
  \caption{
    Storage in number of floating point values w.r.t. number of computational cells/degrees of freedom $N$. In TG methods $r$ denotes the tensor rank and $R$ denotes the number of Sinc-quadrature nodes.
  }
  \label{tab1} 
\end{table}

\begin{table}
  \centering
  \begin{tabular}{l c c c c c}
    Method  &  Setup &  Potential & Field & Energy\\ \hline\hline
    DM          &  $\mathcal{O}(N \log N)$ & --             & $\mathcal{O}(N \log N)$ & $\mathcal{O}(N)$\\ \hline
    SP          &  $\mathcal{O}(N \log N)$ & $\mathcal{O}(N \log N)$  & $\mathcal{O}(N)$ & $\mathcal{O}(N)$\\ \hline
    TG (dense)  &  $\mathcal{O}(N^{2/3}) $ & $\mathcal{O}(N^{4/3})$   & $\mathcal{O}(N)$ & $\mathcal{O}(N)$\\ \hline 
    TG (Tucker) &  $\mathcal{O}(N^{2/3}) $ & $\mathcal{O}(rN^{2/3})$  & $\mathcal{O}(rN^{1/3})$ & $\mathcal{O}(r^2N^{1/3}+r^4)$\\ \hline
    TG (CP)     &  $\mathcal{O}(N^{2/3}) $ & $\mathcal{O}(rN^{2/3})$  & $\mathcal{O}(rN^{1/3})$ & $\mathcal{O}(r^2 N^{1/3})$ \\ \hline
    FES         &  $\mathcal{O}(N)$        & $\mathcal{O}(N log^\alpha N), \alpha \ll 1$   & $\mathcal{O}(N)$ & $\mathcal{O}(N)$\\ \hline
  \end{tabular}
  \caption{
    Computational complexity w.r.t. number of computational cells/degrees of freedom $N$. In TG methods $r$ denotes the tensor rank.
    Every column depends on its left neighbor, e.g. the calculation of the field requires the previous calculation of the potential etc.
  }
  \label{tab2} 
\end{table}

\section{Numerical Experiments}\label{exps}
\begin{figure}
  \centering
  \subfloat[]{\includegraphics[trim=200 0 200 0,clip,width=0.3\textwidth]{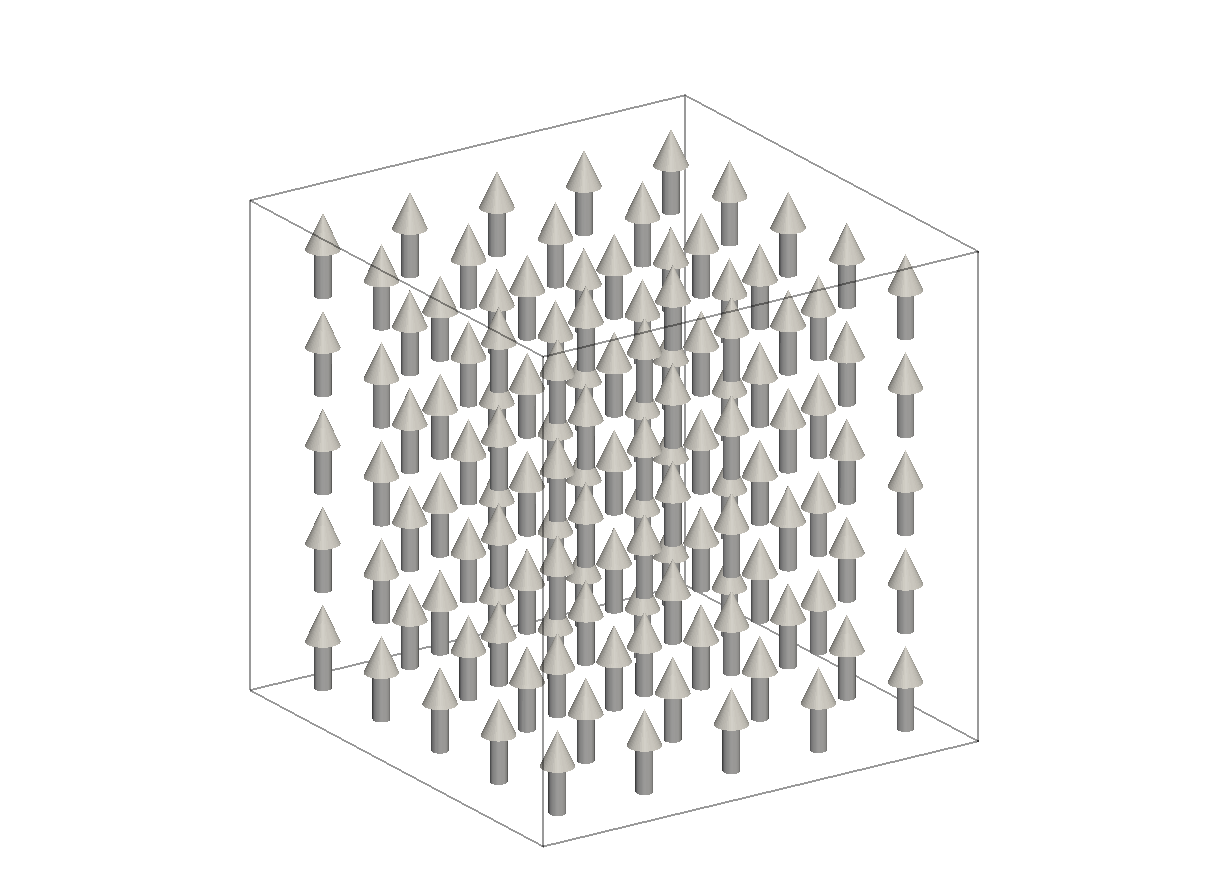}\label{fig:cube_mag_homo}}
  \subfloat[]{\includegraphics[trim=200 0 200 0,clip,width=0.3\textwidth]{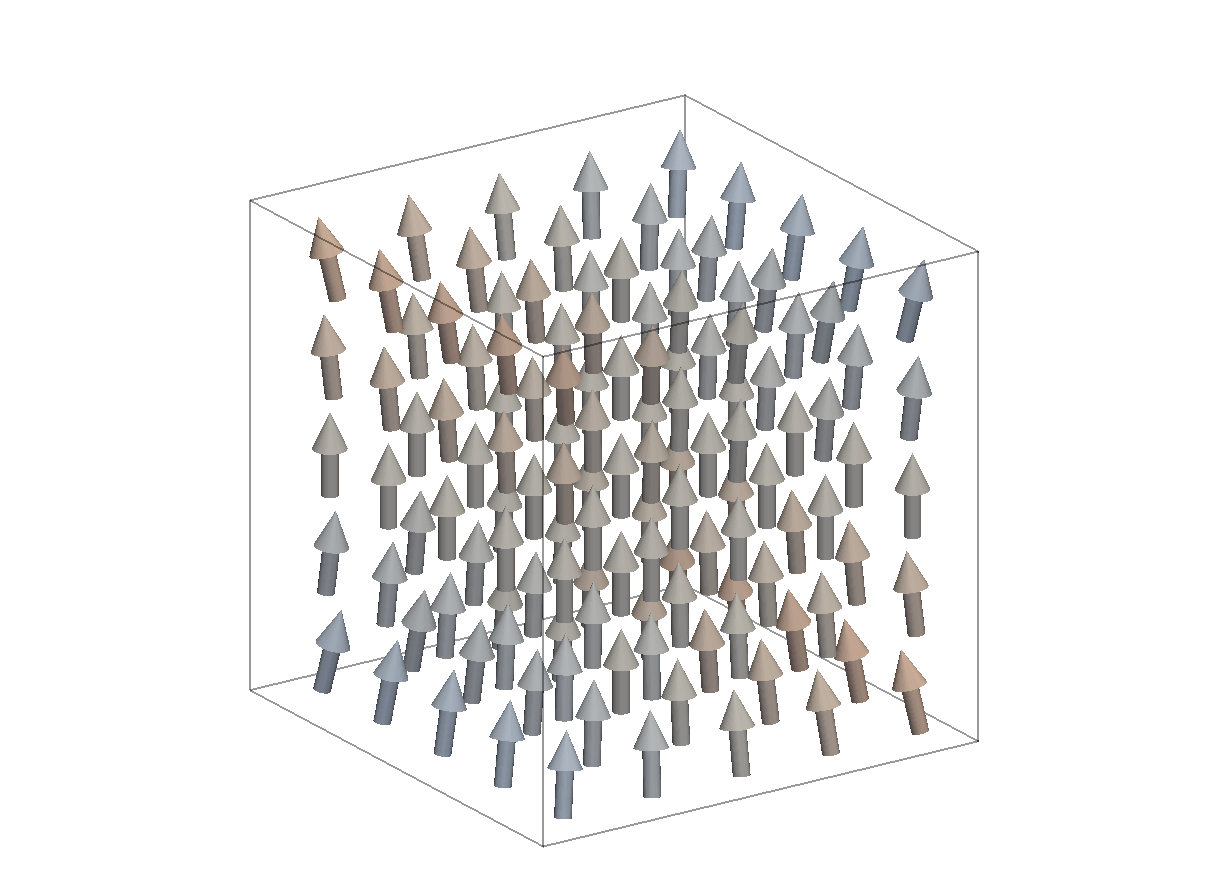}\label{fig:cube_mag_flower}}
  \subfloat[]{\includegraphics[trim=200 0 200 0,clip,width=0.3\textwidth]{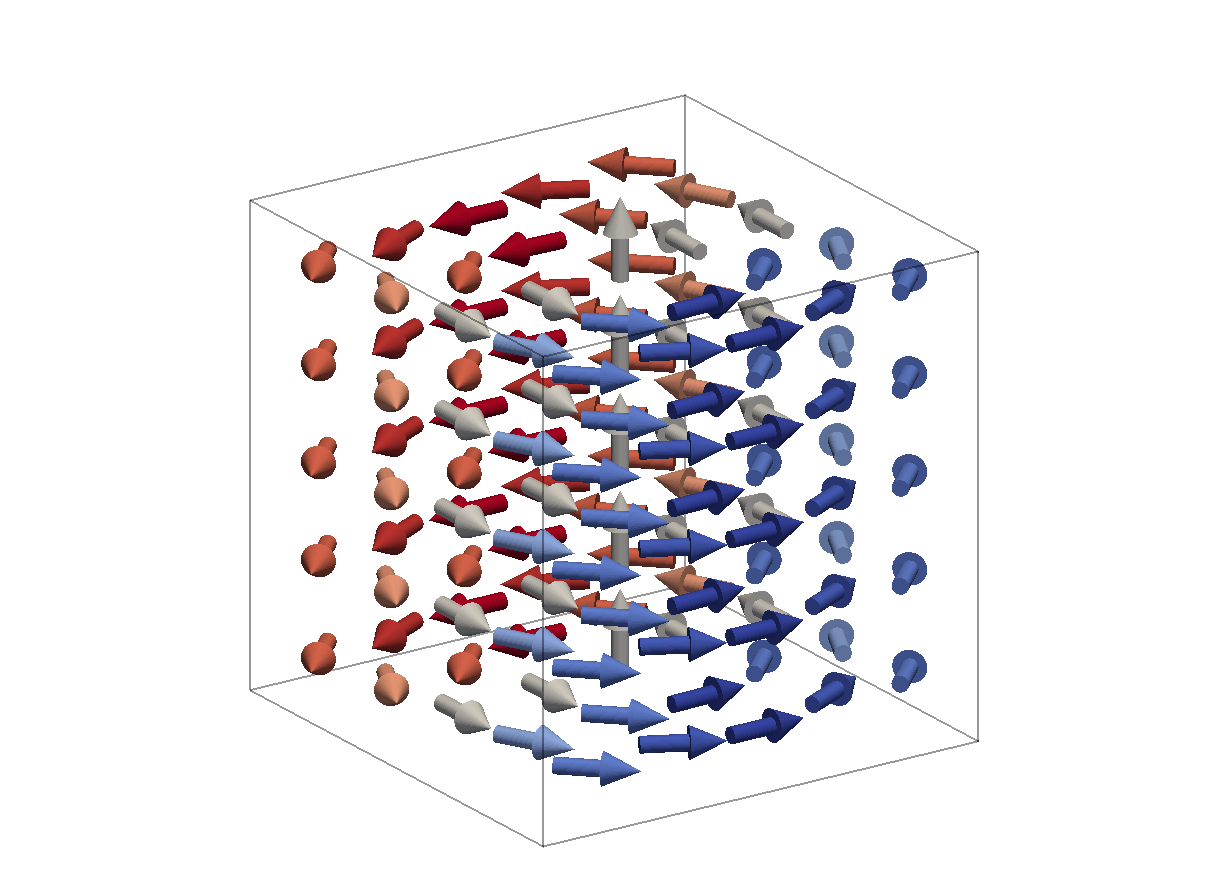}\label{fig:cube_mag_vortex}}
  \caption{
    Magnetization configurations in a $1 \times 1 \times 1$ cube used for numerical experiments.
    The magnetization is normalized, its direction is color coded.
    (a) homogeneous magnetization
    (b) fower state
    (c) vortex state.
  }
  \label{fig:cube_mag}
\end{figure}
\begin{figure}
  \centering
  \subfloat[]{\includegraphics[trim=200 0 200 500,clip,width=0.3\textwidth]{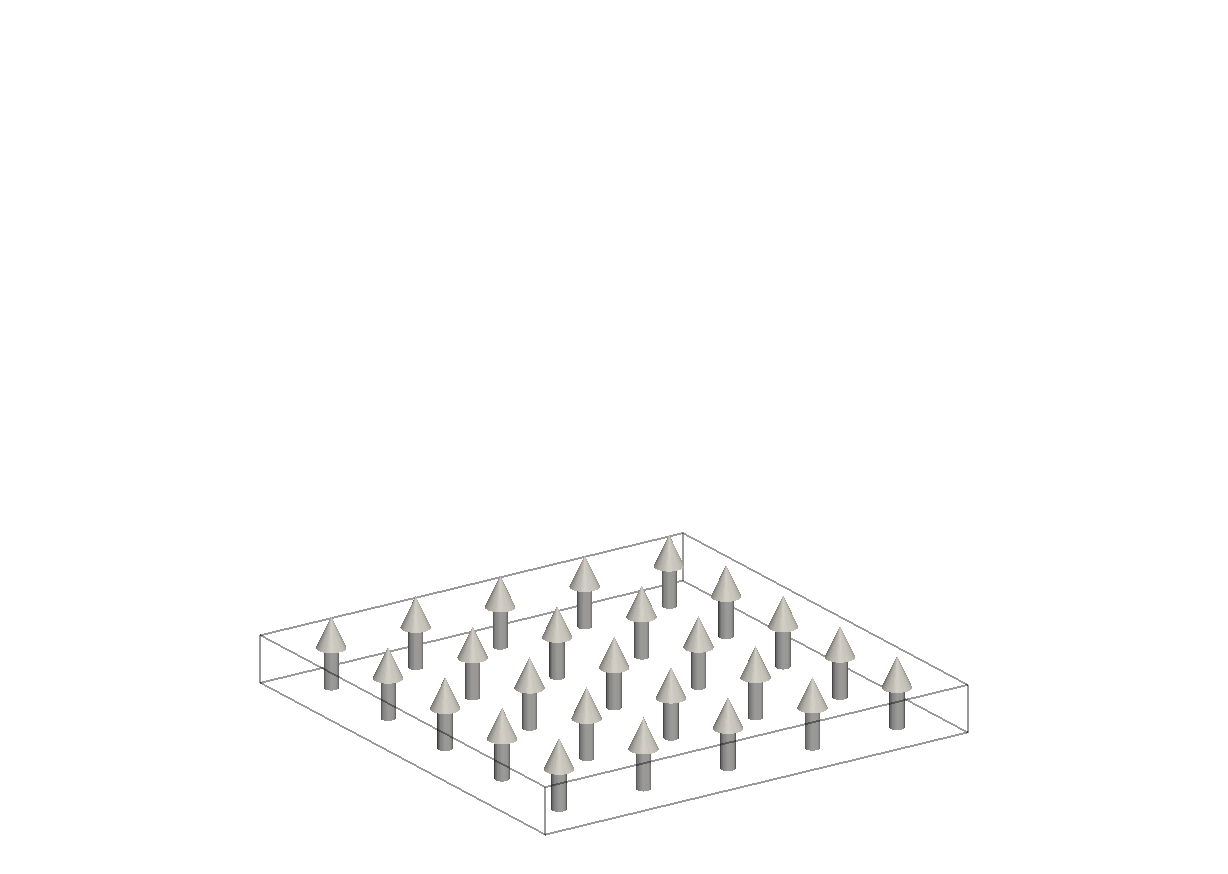}\label{fig:thin_mag_homo}}
  \subfloat[]{\includegraphics[trim=200 0 200 500,clip,width=0.3\textwidth]{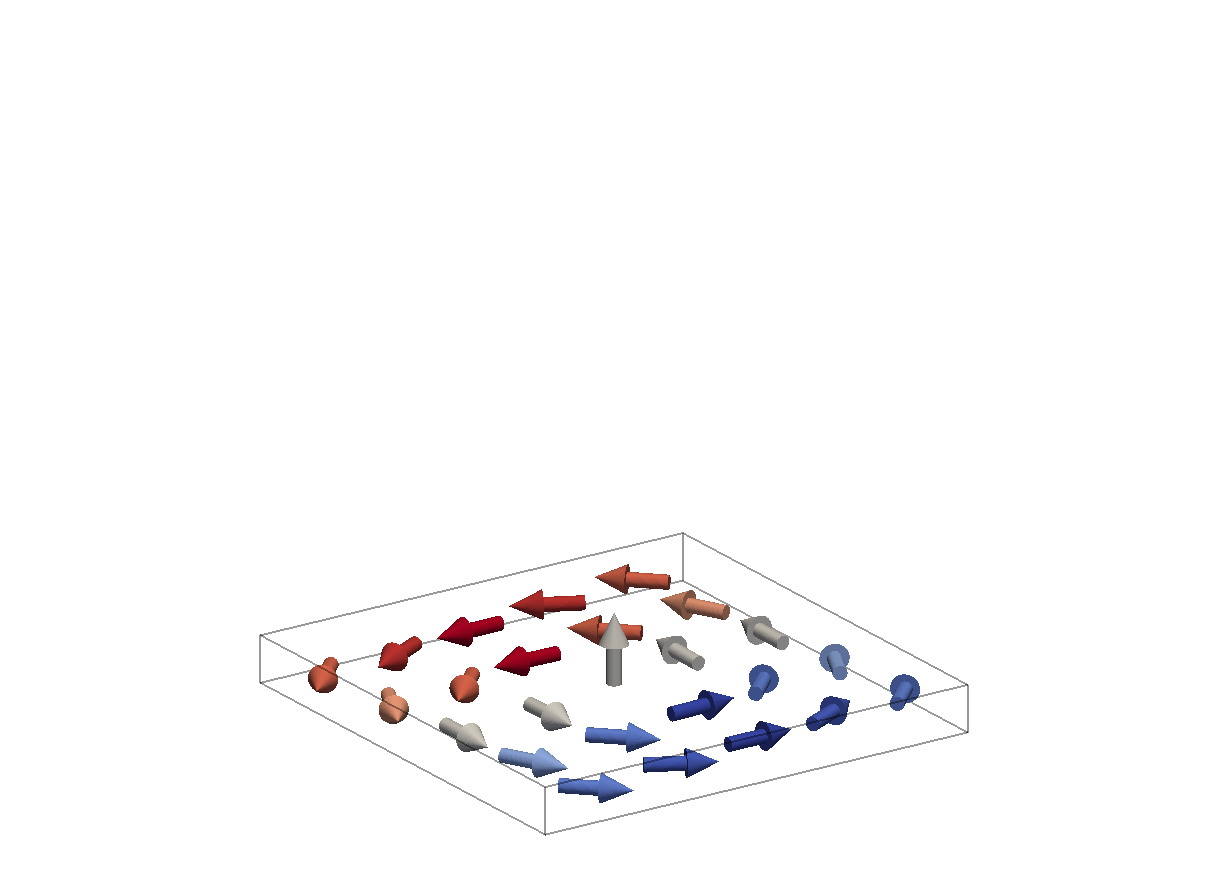}\label{fig:thin_mag_vortex}}
  \caption{
    Magnetization configurations in a $1 \times 1 \times 0.1$ cuboid used for numerical experiments.
    The magnetization is normalized, its direction is color coded.
    (a) homogeneous magnetization
    (b) vortex state.
  }
  \label{fig:thin_mag}
\end{figure}
\begin{figure}
  \centering
  \subfloat[]{\includegraphics[width=0.4\textwidth]{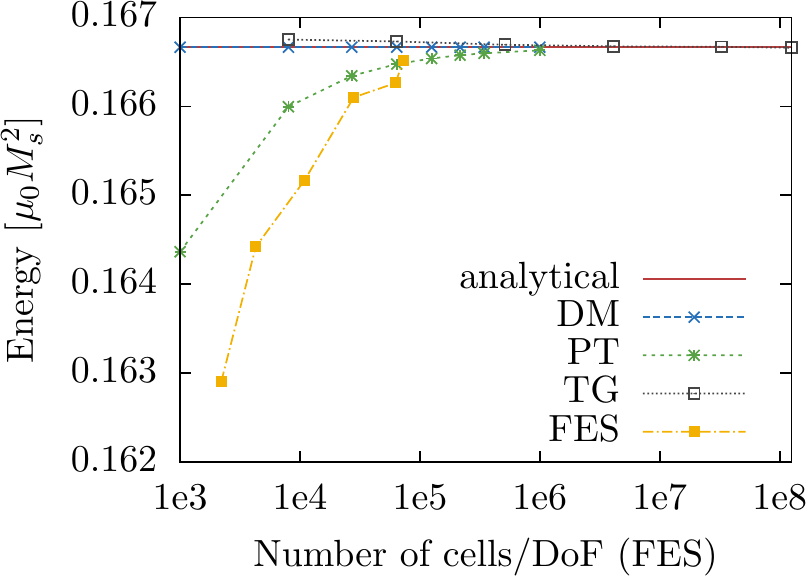}\label{fig:convergence_cube_homo}} \hspace{1cm}
  \subfloat[]{\includegraphics[width=0.4\textwidth]{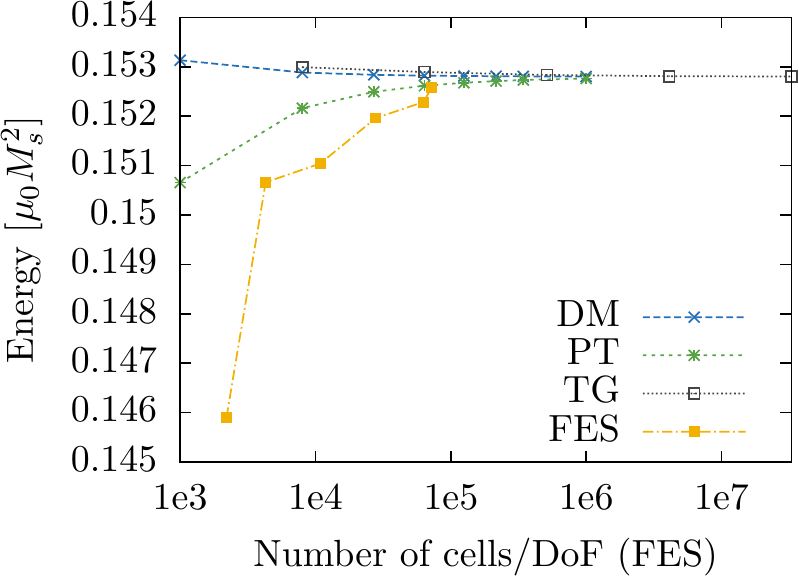}\label{fig:convergence_cube_flower}} \\
  \subfloat[]{\includegraphics[width=0.4\textwidth]{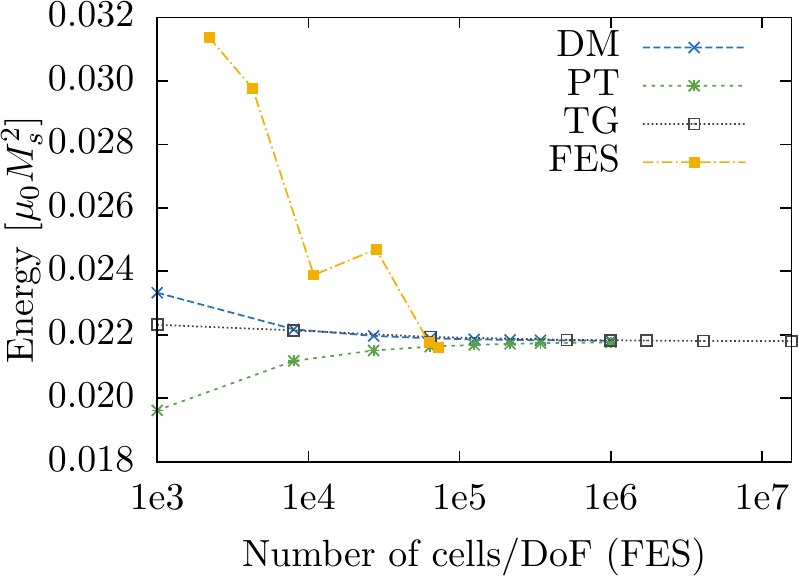}\label{fig:convergence_cube_vortex}} \hspace{1cm}
  \subfloat[]{\includegraphics[width=0.4\textwidth]{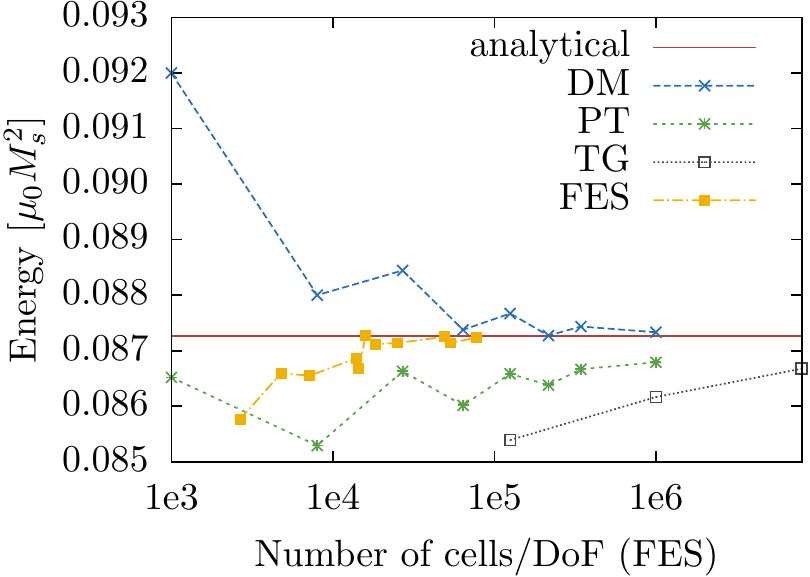}\label{fig:convergence_sphere_homo}}
  \caption{
    Convergence of the calculated stray-field energy for different geometries and magnetization configurations.
    Like in Tab.~\ref{tab1} and \ref{tab2}, $N$ is the number of cells for the tensor grid methods (DM, SP and TG) and the number of degrees of freedom in the case of finite elements (FES).
    (a) homogeneously magnetized $1 \times 1 \times 1$ cube
    (b) flower state in $1 \times 1 \times 1$ cube (c) vortex state in $1 \times 1 \times 1$ state (c) homogeneously magnetized sphere with radius 0.5.
  }
  \label{fig:convergence}
\end{figure}

\subsection{Homogeneously Magnetized Cube}\label{homo_cube}
As a first benchmark we take a homogeneously magnetized unit cube and compute the magnetostatic energy for varying grid-size $N$, where the exact value is $e_d = 1/6\, [\mu_0 M_s^2]$.
Tab.~\ref{cube_homo} shows for each of the described methods the relative errors in the energy w.r.t. the exact value and the relative error in the field computed by \eqref{relerr}, as well as the angular deviation (error in the field-angle) to the field computed by the FD-Demag method, see \eqref{demag}.
We take the relative $l_2 -$ error as a measurement for the field-error, i.e.    

\begin{align}\label{relerr}
  relerr = \big(\frac{1}{N}\sum_{p = x,y,z} \left\|\boldsymbol{H}_{\text{demag}}^{(p)} - \boldsymbol{H}_{\text{method}}^{(p)}\right\|_F^2\big)^{1/2}.
\end{align}  

The errors in the field-angle in Tab.~\ref{cube_homo} mostly occure at the edges of the cube.\\[0.1cm]  

Fig.~\ref{fig:convergence_cube_homo} shows magnetostatic energy calculations for different spatial discretizations.
The DM method is almost exact and does not depend on spatial discretization.
The reason is that the discretized demagnetization tensor is computed assuming homogeneously magnetized computational cells.
Also the resulting stray field is analytically averaged per cell.
Since the energy calculation is bilinear in the magnetization $\boldsymbol{M}$ and the stray field $\boldsymbol{H}$, the error is a pure rounding error.

The FES method is the slowest converging method for this problem.
A possible reason is the large external stray field of this setup.
The numerical integration of the diverging metric tensor $\boldsymbol{g}$ leads to an underestimation of the external space and consequently to an underestimation of the magnetic potential in the sample.
Thus the FES method is particularly sensitive to setups with large external stray fields.

SP and TG methods are also based on the computation of the scalar potential, whereby the field is obtained by finite differences. In \cite{exl_fast_2012} it is shown that the TG method essentially computes the scalar potential exactly for piecewise constant magnetization. The error in the energy in both methods (SP and TG) is mostly caused by numerical approximation of the gradient in the field computation, whereas TG shows the better approximation properties for this problem. In addition to it, TG uses an exact rank$-1$ representation for the uniform magnetization which makes the computation sub-linear (namely $\mathcal{O}(N^{2/3})$) with small scaling factor and allows computations for dozens of millions cells without any problems related to storage and computational cost.\\  

\begin{table}
  \centering
  \begin{tabular}{l c c c c c}
    Method  & $N$ &  relerr $e$&  relerr $h$ & av. relerr $h$ [$^{\circ}$] & max. err $h$ [$^{\circ}$]\\ \hline\hline
    DM    & $40 \times 40 \times 40$ & $2.9e-09$ & $-$  & $-$ & $-$\\ \hline
    SP      & $40 \times 40 \times 40$ & $1.1e-03$ & $1.1e-03$ & $2.3e-05$ & $5.0e+00$\\ \hline
    TG (CP\,$r=1$)     & $40 \times 40 \times 40$ & $3.8e-04$ & $2.3e-03$ & $6.9e-06$ & $2.5e+00$\\ \hline
    FES & $7.2e+04$ & $8.6e-04$ & $2.2e-03$ & $3.2e-05$ & $5.2e+00$\\ \hline
  \end{tabular}
  \caption{
    Homogeneously magnetized unit cube, relative error in the energy, the average relative error in the field/field-angle (w.r.t. DM).
  }
  \label{cube_homo} 
\end{table}
\subsection{Flower and Vortex State in a Cube}  
We do the same comparison as in Sec.~\ref{homo_cube} for the flower state, see \eqref{flower_mag} and Fig.~\ref{fig:cube_mag_flower}, and Tab.~\ref{cube_flower} for the results.
The main magnetization direction is taken to be along the $z$\,-\,axis, and the flower is obtained through an in-plane perturbation along the $y$\,-\,axis and an out-of-plane perturbation along the $x$\,-\,axis.
Assuming polynomial expressions for the perturbations, as in \cite{cowburn_1998}, our flower is the normalized version of
\begin{align}\label{flower_mag}
  m_x(r) = & ~ \tfrac{1}{a} x z, \nonumber \\
  m_y(r) = & ~ \tfrac{1}{c} y z + \tfrac{1}{b^3}\,y^3 z^3,  \\
  m_z(r) = & ~ 1 \nonumber,
\end{align}
where the center of the cube is located at $(0, 0, 0)$.
We choose $a=c=1$ and $b=2$.\\[0.2cm] 
\begin{table}
  \centering
  \begin{tabular}{l c c c c c}
    Method               & $N$                      & $e$         & relerr $h$ & av. relerr $h$ [$^{\circ}$] & max. err $h$ [$^{\circ}$]\\ \hline\hline
    DM                   & $40 \times 40 \times 40$ & $1.528e-01$ & $-$  & $-$                  & $-$\\ \hline
    SP                   & $40 \times 40 \times 40$ & $1.526e-01$ & $1.8e-03$                   & $5.0e-05$ & $7.2e+00$\\ \hline
    TG (CP,\,$r=6$)      & $40 \times 40 \times 40$ & $1.529e-01$ & $1.8e-03$                   & $7.8e-06$ & $2.6e+00$\\ \hline
    FES                  & $7.2e+04$ & $1.526e-01$              & $2.5e-03$   & $6.0e-05$                   & $6.8e+00$\\ \hline
  \end{tabular}
  \caption{
    Flower state for magnetization in the unit cube, energy, the average relative error in the field/field-angle (w.r.t. DM).
  }
  \label{cube_flower} 
\end{table}
The results are similar to those of the homogeneously magnetized sample.
In contrast the results of the DM method are not exact in this case, but Fig.~\ref{fig:convergence_cube_flower} shows that the DM method converges faster than all other methods.

The next comparison is for a vortex state in a unit cube, see Fig.~\ref{fig:cube_mag_vortex}, described by the model in \cite{Feldtkeller1965}, i.e. 
\begin{align} \label{config1}
 m_{x}(r) = & ~-\frac{y}{r}\,\big( 1 - \exp\big( -4\,\frac{r^2}{r^2_c} \big) \big)^{\frac{1}{2}}, \nonumber \\
 m_{y}(r) = & ~~\frac{x}{r}\,\big( 1 - \exp\big( -4\,\frac{r^2}{r^2_c} \big) \big)^{\frac{1}{2}},  \\
 m_{z}(r) = & ~\exp\big( -2\,\frac{r^2}{r^2_c} \big) \nonumber,
\end{align}
where $r = \sqrt{x^2 + y^2}$, and we choose the radius of the vortex core as $r_c = 0.14$.
The vortex center coincides with the center of the cube, and the magnetization is assumed to be rotationally symmetric about the $x/y$\,-\,axis
and translationally invariant along the $z$\,-\,axis.
The results can be found in Tab.~\ref{cube_vortex}.

The most notable difference to the previous tests is the large field error in the FES method.
It shows that the error occurs in the center of the vortex, where the gradient of the magnetization peaks.
A possible solution for this problem would be an adaptive meshing, which is currenty not implemented in our FES code.

\begin{table}
  \centering
  \begin{tabular}{l c c c c c}
    Method              & $N$                      & $e$         &  relerr $h$ & av. relerr $h$ [$^{\circ}$] & max. err $h$ [$^{\circ}$]\\ \hline\hline
    DM                  & $40 \times 40 \times 40$ & $2.189e-02$ & $-$         & $-$                         & $-$\\ \hline
    SP                  & $40 \times 40 \times 40$ & $2.163e-02$ & $2.4e-03$   & $3.4e-04$                   & $1.2e+01$\\ \hline
    TG (Tucker,\,$r=10$)& $40 \times 40 \times 40$ & $2.193e-02$ & $4.0e-03$   & $3.2e-04$                   & $1.1e+01$\\ \hline
    FES            & $7.2e+04$ & $2.160e-02$  & $2.1e-02$   & $6.1e-02$                   & $1.7e+02$\\ \hline
  \end{tabular}
  \caption{
    Vortex state for magnetization in the unit cube, energy, the average relative error in the field/field-angle (w.r.t. DM).
  }
  \label{cube_vortex} 
\end{table}

\subsection{Thin Film}
We first take a homogeneously magnetized $1\times 1\times0.1$ thin film (magnetization out of plane), see Tab.~\ref{thin_homo} for the results. Tab.~\ref{thin_vortex} shows the results for the vortex state (out of plane) in the same thin film geometry.\\ 

The results for methods that do not rely on spatial discretization outside the sample perform equally well on this geometry. FES, instead, shows a deterioration of performance due to the worse ratio of shell and sample elements while leaving the number of DoF unchanged.
\begin{table}
  \centering
  \begin{tabular}{l c c c c c}
    Method          & $N$                  & $e$         &  relerr $h$ & av. relerr $h$ [$^{\circ}$] & max. err $h$ [$^{\circ}$]\\ \hline\hline
    DM              & $80\times 80\times8$ & $4.025e-02$ & $-$         & $-$                         & $-$\\ \hline
    SP              & $80\times 80\times8$ & $4.021e-02$ & $1.7e-03$   & $2.6e-05$                   & $4.5e+00$\\ \hline
    TG (CP,\,$r=1$) & $80\times 80\times8$ & $4.025e-02$ & $3.7e-03$   & $6.4e-06$                   & $2.1e+00$\\ \hline
    FES        & $4.9e+04$            & $3.983e-02$ & $5.5e-03$   & $1.9e-05$                   & $5.0e+00$\\ \hline
    \end{tabular}
  \caption{
    Homogeneously magnetized $1\times1\times0.1$ thin film (magnetization out of plane), energy, the average relative error in the field/field-angle (w.r.t. DM).
  }
  \label{thin_homo} 
\end{table}

\begin{table}
  \centering
  \begin{tabular}{l c c c c c}
    Method          & $N$                  & $e$         &  relerr $h$ & av. relerr $h$ [$^{\circ}$] & max. err $h$ [$^{\circ}$]\\ \hline\hline
    DM              & $80\times 80\times8$ & $1.569e-03$ & $-$         & $-$                         & $-$\\ \hline
    SP              & $80\times 80\times8$ & $1.555e-03$ & $2.4e-03$   & $4.6e-05$                   & $3.6e+00$\\ \hline
    TG (CP,\,$r=1$) & $80\times 80\times8$ & $1.569e-03$ & $2.9e-03$   & $1.8e-05$                   & $4.0e+00$\\ \hline
    FES        & $4.9e+04$            & $1.496e-03$ & $6.0e-03$         & $6.4e-04$                         & $2.1e+01$\\ \hline
  \end{tabular}
  \caption{
    Vortex magnetization in $1\times1\times0.1$ thin film (magnetization out of plane), energy, the average relative error in the field/field-angle (w.r.t. DM).
  }
  \label{thin_vortex} 
\end{table}
\subsection{Sphere}
\begin{figure}
  \centering
  \subfloat[]{\includegraphics[trim=220 80 220 80,clip,width=0.3\textwidth]{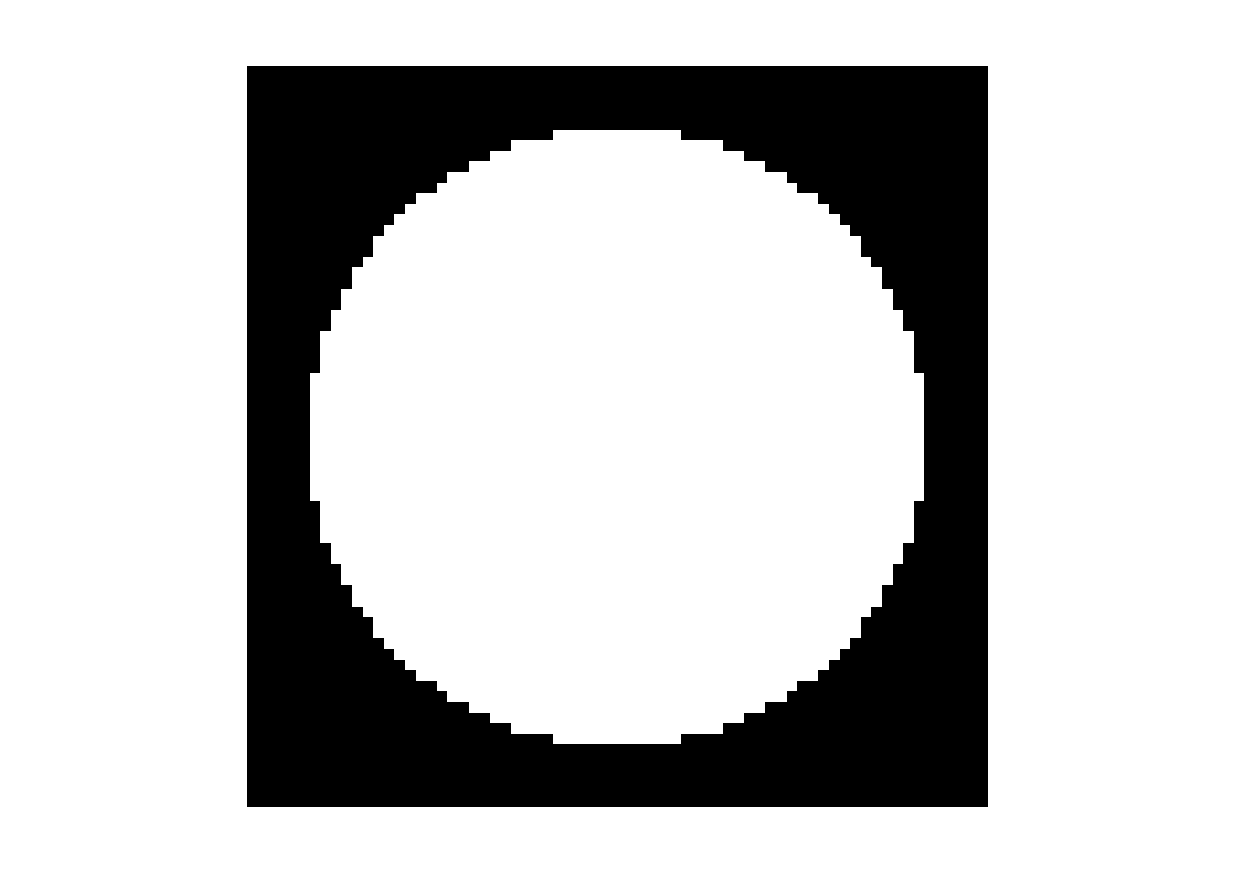}\label{fig:sphere_mag_fdm}}
  \subfloat[]{\includegraphics[trim=220 80 220 80,clip,width=0.3\textwidth]{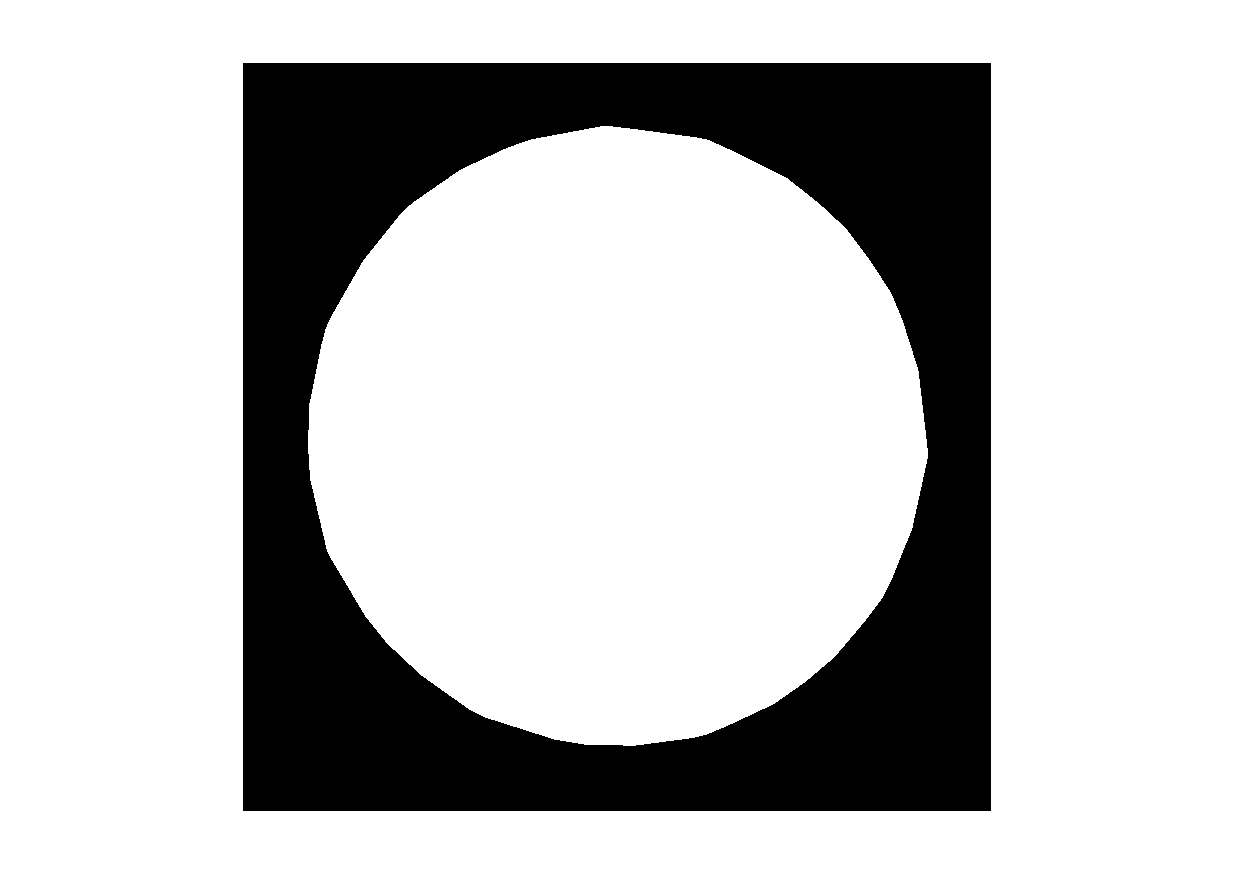}\label{fig:sphere_mag_fem}}\\
  \subfloat[]{\includegraphics[trim=220 80 220 80,clip,width=0.3\textwidth]{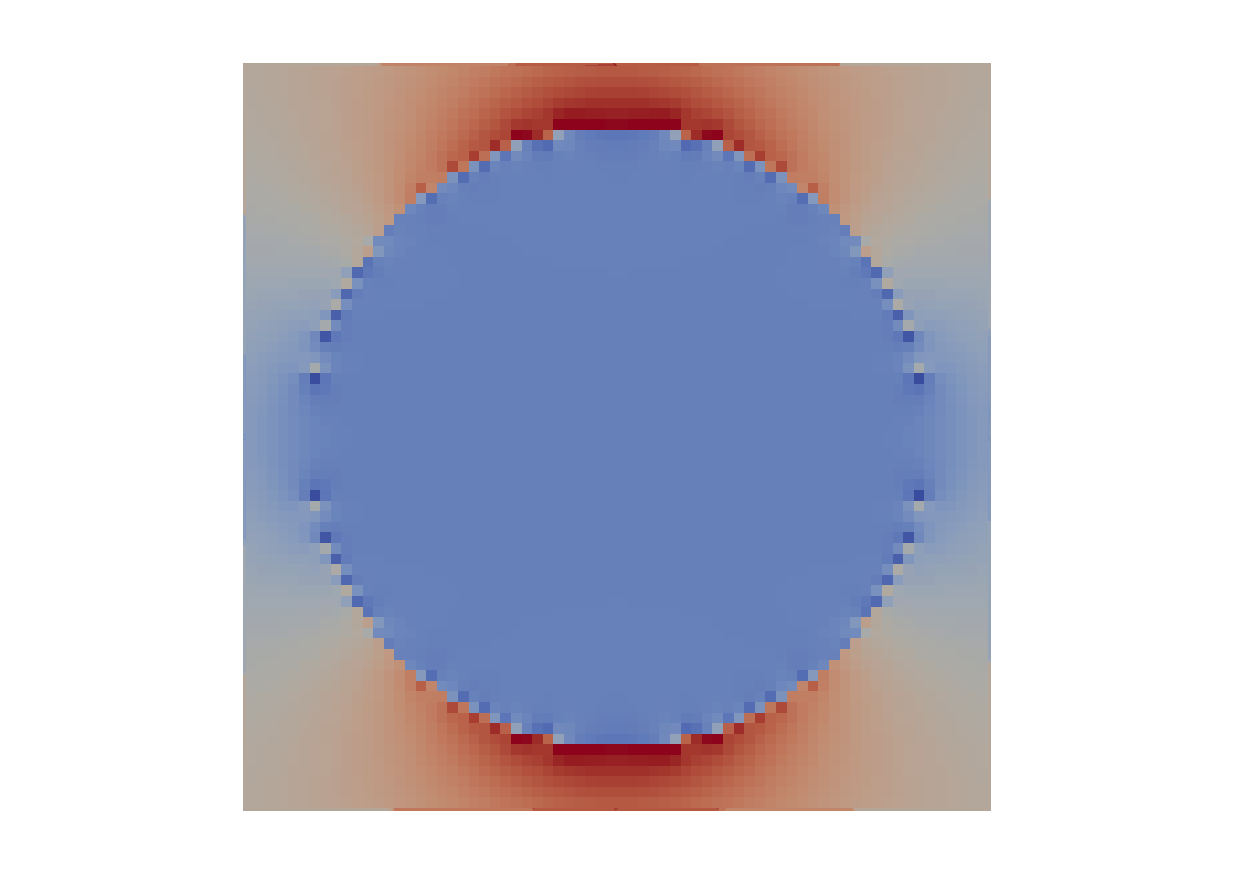}\label{fig:sphere_h_fdm}}
  \subfloat[]{\includegraphics[trim=220 80 220 80,clip,width=0.3\textwidth]{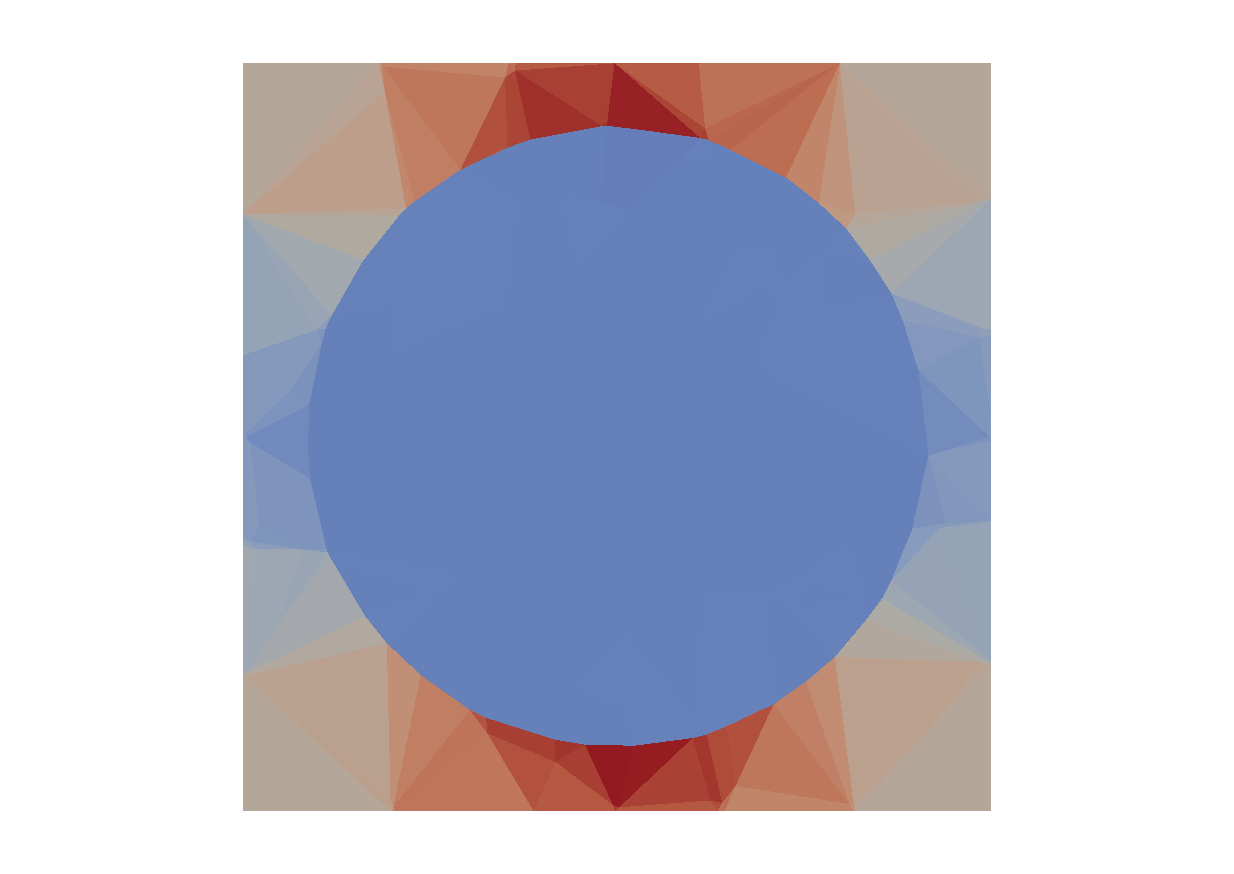}\label{fig:sphere_h_fem}}
  \caption{
    Spatial discretization and stray field of a homogeneously, in $z$-direction magnetized sphere in the middle $xz$-plane
    (a) Finite difference approximation of the spherical sample with $50 \times 50 \times 50$ cells
    (b) Finite element approximation of the spherical sample with 9429 tetrahedra (including the shell elements)
    (c) $z$ component of the stray field, calculated with DM method
    (d) $z$ component of the stray field, calcualted with FES method
  }
\end{figure}
As the last test a homogeneously magnetized sphere with radius $R = 0.5$ is simulated.
The spatial discretization in case of cuboid grids is done by setting the magnetization $\boldsymbol{M} = (0, 0, M_z)$ in cells whose center lies within the sphere.
This leads to staircase artifacts as shown in Fig.~\ref{fig:sphere_mag_fdm}.
For the FES method the sphere is discretized such that the volume of the discretized sphere matches the anaytical volume.

The magnetostatic energy for different spatial discretizations is displayed in Fig.~\ref{fig:convergence_sphere_homo}.
The FES method shows the fastest convergence, which is obviously a consequence of the better approximation of the curved surface, see Fig.~\ref{fig:sphere_mag_fem}.
Also the field computation benefits from this better approximation, see Fig.~\ref{fig:sphere_h_fdm} and \ref{fig:sphere_h_fem}.

Work was done on the treatment of curved surfaces within cartesian grid methods \cite{donahue_2007}.
Still the use of irregular grids is a more natural way of describing curved surfaces and is thus preferable.

\section{Conclusion}
We investigated several test magnetization configurations with different methods for the stray-field computation and compared the results.
There is no clear winner in this comparison of numerical methods for the stray-field calculation.
Computations on cuboid structures are best done with methods that compute on cuboid grids, namely the DM, SP and TG methods.
The TG method is not only the fastest choice, it is also able to handle very large grids due to low-rank tensor approximation or representation of the magnetization.
However the TG method is not yet well investigated in the context of full micromagnetic simulations.
In order to preserve the sublinear complexity and storage requirement features further research on the behaviour of low-rank magnetization during energy minimization or LLG integration has to be done.

The SP method is faster than the DM method by a factor of 1.5 and needs about 30\% less memory.
This speedup comes at the expense of accuracy.
Among the Cartesian grid methods, the DM method is most accurate since the stray field is computed directly. 
Both the SP and the TG method show an additional error due to the finite-difference gradient computation.

For curved structures FES is a good choice.
The obvious reason for this is the use of irregular meshes, which are able to model the curvature much better than cuboid grids. In contrast to FEM/BEM methods FES can be implemented using sparse matrices only, since the presence of the dense boundary element matrix is overcome with the shell transformation.

\section*{Acknowledgements}
The authors want to thank Michael Hinze and Winfried Auzinger for fruitful discussions.
Financial support by
the Deutsche Forschungsgemeinschaft via the Graduiertenkolleg 1286 {\emph ``Functional Metal-Semiconductor Hybrid Systems''},
the Austrian Science Fund (FWF) SFB ViCoM (F4112-N13),
and the Sonderforschungsbereich 668 ``Magnetism from the single atom to the nanostructure''
is gratefully acknowledged.

\appendix
\section{Low-Rank Tensor Formats}\label{low-rank}\label{appendix} 
A tensor $\boldsymbol{\mathcal{A}}\in\mathbb{R}^{N_1 \times N_2 \times N_3}$ is said to be in
\textit{canonical format (CANDECOMP/PARAFAC (CP) decomposition)} with \textit{tensor product rank}\, $ r$, if
\begin{align} \label{CP}
\boldsymbol{\mathcal{A}} = \sum_{l=1}^{r} \lambda_l\,\,\boldsymbol{u}^{(1)}_l \circ \boldsymbol{u}^{(2)}_l \circ \boldsymbol{u}^{(3)}_l
\end{align}
with $\lambda_l \in \mathbb{R} $,\, 
 vectors $\boldsymbol{u}^{(j)}_l \in \mathbb{R}^{N_j}$, 
and $\circ$ is the vector outer product. A particular entry of a canonical tensor is given by
\begin{align}
a_{ijk} = \sum_{l=1}^{r} \lambda_l\,\big(u^{(1)}_l\big)_i \, \big(u^{(2)}_l\big)_j \, \big(u^{(3)}_l\big)_k .
\end{align}

Abbreviating notation as in \cite{kolda_tensor_2009}, a tensor $\boldsymbol{\mathcal{A}}\in \mathbb{R}^{N_1 \times N_2 \times N_3}$ in CP format can be written as
\begin{align}\label{CP_mat}
\boldsymbol{\mathcal{A}} = \llbracket \boldsymbol{\lambda};\,\boldsymbol{U}^{(1)},\boldsymbol{U}^{(2)},\boldsymbol{U}^{(3)} \rrbracket,
\end{align}
with weight vector $\boldsymbol{\lambda} = [ \lambda_1,\ldots,\lambda_r ] \in \mathbb{R}^{r}$
and factor matrices $\boldsymbol{U}^{(j)} = \big[\,\boldsymbol{u}^{(j)}_1 \,|\, \hdots \,|\, \boldsymbol{u}^{(j)}_r\,\big] \in \mathbb{R}^{N_j \times r}$.\\ 
From \eqref{CP_mat} it can be seen that the number of degrees of freedom (DoF) of a CP tensor is $r+r\sum_j N_j$ (compare with $\prod_j N_j$ for a dense tensor), also see Tab.~\ref{tab1}.\\[0.1cm]

  
A tensor $ \boldsymbol{\mathcal{A}}\in \mathbb{R}^{N_1 \times N_2 \times N_3} $ is said to be in \textit{Tucker format} (Tucker tensor) if it can be represented in the form
\begin{align} \label{tucker}
\boldsymbol{\mathcal{A}} = \boldsymbol{\mathcal{C}} \times_1 \boldsymbol{U}_1 \times_2 \boldsymbol{U}_2 \times_3 \boldsymbol{U}_3,
\end{align}
with the so-called \textit{core tensor} $\boldsymbol{\mathcal{C}} \in \mathbb{R}^{r_1 \times r_2 \times r_3}$
and \textit{factor matrices} $\boldsymbol{U}_j \in \mathbb{R}^{N_j \times r_j}$.\\
Hereby the key-operation is the $n$-mode (matrix) multiplication of a tensor $\boldsymbol{\mathcal{A}}\in\mathbb{R}^{N_1 \times N_2 \times N_3}$ with a matrix $\boldsymbol{U}\in\mathbb{R}^{M\times N_n}$, which is the multiplication 
of each mode-$n$ fiber of $\boldsymbol{\mathcal{A}}$ by the matrix $\boldsymbol{U}$, i.e.
\begin{align}
\boldsymbol{\mathcal{A}} \times_{n} \boldsymbol{U} \in \mathbb{R}^{\times_{j=1}^{3} M_j},\,\,\, M_j = 
\left\{\begin{array}{c}
N_j , \, j \neq n\\
M , \, j = n.
\end{array}\right.
\end{align}

In contrast to CP tensors, the ranks in the Tucker representation can be different in each mode (dimension). In the discussions of Sec.~\ref{storage} and the experiments in Sec.~\ref{exps} we used the same rank $r$ for each mode, i.e. $r \equiv r_j$.\\
For a tensor in Tucker format $\prod_j r_j + \sum_j r_j\, N_j$ entries have to be stored, which is a compression for $r_j \ll N_j$, also see Tab.~\ref{tab1}.\\

For a sum of Tucker tensors one can only store the factor matrices and core tensors of the summands, which is called \textit{block-CP format}.\\[0.2cm]

Linear algebra operations for low-rank tensors, like the inner product, tensor--matrix product etc., can be performed without forming the dense tensors \cite{bader_efficient_2008}, which makes this operations faster than their conventional counterparts. 

\bibliographystyle{plainnat}
\bibliography{refs}

\end{document}